\documentclass[prl,aps,amsfonts,amssymb,showpacs,preprint]{revtex4}

\usepackage{graphicx,bm}

\def\norm#1{\left|\mkern-2mu\left|#1\right|\mkern-2mu\right|}

\begin{document}

\title{Enstrophy dissipation in freely evolving two-dimensional turbulence}

\author{Chuong V. Tran}
\email{ctran@maths.warwick.ac.uk}

\affiliation{Mathematics Institute, University of Warwick, 
Coventry CV4 7AL, UK}

\date{\today}

\begin{abstract}

Freely decaying two-dimensional Navier--Stokes turbulence is studied. The 
conservation of vorticity by advective nonlinearities renders a class of  
Casimirs that decays under viscous effects. A rigorous constraint on the 
palinstrophy production by nonlinear transfer is derived, and an upper 
bound for the enstrophy dissipation is obtained. This bound depends only 
on the decaying Casimirs, thus allowing the enstrophy dissipation to be 
bounded from above in terms of initial data of the flows. An upper bound 
for the enstrophy dissipation wavenumber is derived and the new result is
compared with the classical dissipation wavenumber. 
     
\end{abstract}


\maketitle

In 1969, Batchelor$^1$ adapted Kolmogorov's equilibrium theory for 
three-dimensional (3D) turbulence to two-dimensional (2D) turbulence, on 
the basis of a phenomenologically analogous property between the two 
systems. For a 3D fluid, decrease of the viscosity $\nu$ is accompanied 
by increase of the mean-square vorticity, a consequence of the magnification 
of the vorticity by stretching of vortex lines, so that in the inviscid 
limit the energy dissipation is nonzero. For a 2D fluid, decrease of the 
viscosity enhances convective mixing, in which isovorticity lines get 
extended and brought closer to one another,$^1$ giving rise to increase 
of the mean-square vorticity gradients (twice the palinstrophy), so that 
in the inviscid limit the rate of enstrophy (half the mean-square vorticity) 
dissipation can approach a finite value $\chi$. On the basis of this analogy, 
Batchelor$^1$ applies the familiar arguments of the Kolmogorov equilibrium 
theory for the small-scale components of 3D turbulence to the 2D case, 
where the roles of the energy and energy dissipation in the original 
theory are played by the enstrophy and enstrophy dissipation. This means 
that the statistical properties of the small-scale components of the 
turbulence depend only on the two dimensional parameters $\chi$ and $\nu$. 
The enstrophy dissipation $\chi$ is thus an important dynamical quantity 
in Batchelor's theory. One of its prominent role is in the expression of 
the enstrophy spectrum $Z(k)$ of the so-called enstrophy inertial range, 
which is presumably formed when an initial enstrophy reservoir spreads 
out in a virtually inviscid region of wavenumber space:
\begin{eqnarray}
\label{Zspectrum}
Z(k) &=& C\chi^{2/3}k^{-1},
\end{eqnarray}
where $C$ is a universal constant and $k$ is the wavenumber. Another important
role of $\chi$ is in the determination of the dissipation wavenumber $k_\nu$:
\begin{eqnarray}
\label{knu}
k_\nu &=& \frac{\chi^{1/6}}{\nu^{1/2}},
\end{eqnarray}
which presumably marks the end of the enstrophy inertial range, around
which the enstrophy is most strongly dissipated. In both (\ref{Zspectrum}) 
and (\ref{knu}), $\chi$ is a finite (but otherwise undetermined) parameter. 

It is desirable to have a quantitative knowledge of $\chi$ (and hence 
of $k_\nu$), not only for its role in the Batchelor theory but also 
for further analyses of the turbulence, beyond the usual dimensional 
arguments. Even for finite Reynolds numbers, the determination of $\chi$ 
is highly non-trivial. In the limit $\nu\rightarrow0$, this problem does 
not seem to become more tractable. Ideally, $\chi$ can be determined if 
the extent to which the production of palinstrophy by convective mixing is 
fully understood. This letter takes a direct approach to this problem by 
deriving a rigorous upper bound for the nonlinear term representing the 
palinstrophy production rate. Equating this bound to the viscous 
dissipation term yields a constraint, from which upper bounds for $\chi$ 
and for the enstrophy dissipation wavenumber $k_d$ (to be defined later 
in this letter) can be derived. These bounds are found to be completely 
described in terms of initial data of the flows. The derived enstrophy 
dissipation wavenumber is consistent with the classical prediction of 
$k_\nu$ given by (\ref{knu}), in the sense that they both have the same 
functional dependence on $\nu$. A novel result of this study is that 
$\chi$ and $k_d$ can be estimated in terms of initial data of the 
turbulence, while they (more precisely $\chi$ and $k_\nu$) are essentially 
undetermined in the Batchelor theory.

In the vorticity formulation, the freely evolving 2D Navier--Stokes 
equations governing the motion of an incompressible fluid confined to a 
doubly periodic domain are
\begin{eqnarray}
\label{NS}
\partial_t\xi+J(\psi,\xi) &=& \nu\Delta\xi,
\end{eqnarray}
where $\xi(\bm x,t)$ is the vorticity, 
$J(\theta,\vartheta)=\theta_x\vartheta_y-\theta_y\vartheta_x$, $\nu$ 
the kinematic viscosity, and $\psi(\bm x,t)$ the stream function. 
The vorticity is defined in terms of the stream function and of the 
velocity $\bm v$ by $\xi=\Delta\psi=\hat n\cdot\nabla\times\bm v$, 
where $\hat n$ is the normal vector to the fluid domain. 
Equivalently, $\bm v$ can be recovered from $\psi$ and $\xi$ by
$\bm v=(-\psi_y,\psi_x)=(-\Delta^{-1}\xi_y,\Delta^{-1}\xi_x)$.

An importance property of the advective nonlinear term in (\ref{NS}) is 
that it conserves the kinetic energy and an infinite class of integrated
quantities, known as Casimirs, including the enstrophy. The latter 
conservation law is attributed to the fact that vorticity is conservatively 
redistributed in physical space by the advective transfer. While the 
conservation of energy and enstrophy imposes strict constraints on 
turbulent flows and has been explored in the literature to a great 
extent,$^{2-19}$ the conservation of Casimirs, other than the enstrophy, 
seems to render little additional knowledge of the flows and has received 
much less attention.$^{20-22}$ In the present case of unforced dynamics, 
it is well known that a wide class of these Casimirs decays under the 
action of viscosity. Here our main interest is in the following Casimirs: 
$\langle|\xi|^p\rangle$, where $\langle\cdot\rangle$ denotes a spatial 
average. By taking the time derivative of $\langle|\xi|^p\rangle$ and 
using (\ref{NS}) one obtains 
\begin{eqnarray}
\label{Cevolution1}
\frac{d}{dt}\langle|\xi|^p\rangle &=& 
-p\langle|\xi|^{p-2}\xi J(\psi,\xi)\rangle 
+ \nu p\langle|\xi|^{p-2}\xi\Delta\xi\rangle \nonumber\\
&=& -\nu p(p-1)\langle|\xi|^{p-2}|\nabla\xi|^2\rangle,
\end{eqnarray}
where the second equation is obtained by integration by parts, upon which 
the nonlinear term identically vanishes. For $p=2$, Eq. (\ref{Cevolution1}) 
governs the decay of $\langle|\xi|^2\rangle$ (twice the enstrophy), for which 
the dissipation term becomes $2\chi$, which is the subject of this study. 
The right-hand side of (\ref{Cevolution1}) is negative for $p>1$. Hence 
$\langle|\xi|^p\rangle$ decays in time, for $p>1$. It follows that
\begin{eqnarray}
\label{maximum}
\langle|\xi(t)|^p\rangle &\le& \langle|\xi(0)|^p\rangle,
\end{eqnarray}
for $p>1$ and $t\ge0$. In particular, in the limiting case 
$p\rightarrow\infty$, one has
\begin{eqnarray}
\label{maximum1}
\norm{\xi(t)}_\infty &\le& \norm{\xi(0)}_\infty,
\end{eqnarray}
for $t\ge0$, where $\norm{\xi}_\infty$ denotes the $L^\infty$ norm of $\xi$.

Now the main result of this letter can be readily derived. By multiplying 
(\ref{NS}) by $\Delta\xi$ and taking the spatial average of the resulting 
equation, one obtains the equation governing the evolution of the 
palinstrophy $\langle|\nabla\xi|^2\rangle/2$:
\begin{eqnarray}
\label{Pevolution}
\frac{1}{2}\frac{d}{dt}\langle|\nabla\xi|^2\rangle &=& 
\langle\Delta\xi J(\psi,\xi)\rangle - \nu\langle|\Delta\xi|^2\rangle 
\nonumber\\
&=& \langle\xi J(\psi_x,\xi_x)\rangle + \langle\xi J(\psi_y,\xi_y)\rangle 
- \nu\langle|\Delta\xi|^2\rangle \nonumber\\
&\le& \langle|\xi|(|\nabla\psi_x||\nabla\xi_x|+
|\nabla\psi_y||\nabla\xi_y|)\rangle - \nu \langle|\Delta\xi|^2\rangle
\nonumber\\
&\le& \norm\xi_\infty\langle|\nabla\psi_x|^2+|\nabla\psi_y|^2\rangle^{1/2} 
\langle|\nabla\xi_x|^2+|\nabla\xi_y|^2\rangle^{1/2} 
- \nu\langle|\Delta\xi|^2\rangle \nonumber\\
&\le& \norm\xi_\infty\langle|\xi|^2\rangle^{1/2}
\langle|\Delta\xi|^2\rangle^{1/2}  - \nu\langle|\Delta\xi|^2\rangle.
\end{eqnarray} 
In (\ref{Pevolution}), the second equation is obtained via the two 
elementary identities
\begin{eqnarray}
\Delta J(\psi,\xi) &=& J(\psi,\Delta\xi)+2J(\psi_x,\xi_x)+2J(\psi_y,\xi_y)
\end{eqnarray}
and
\begin{eqnarray}
\langle\Delta\xi J(\psi,\xi)\rangle &=& -\langle\xi J(\psi,\Delta\xi)\rangle.
\end{eqnarray}
The H\"{o}lder inequality is used in the fourth step, and the last step 
can be seen by expressing $\psi$ (and $\xi$) in terms of Fourier series. 
The triple-product term $\norm\xi_\infty\langle|\xi|^2\rangle^{1/2}
\langle|\Delta\xi|^2\rangle^{1/2}$ 
represents an upper bound for the palinstrophy production rate. It can be 
seen that the palinstrophy necessarily ceases to grow as its dissipation 
$\nu\langle|\Delta\xi|^2\rangle$ reaches this bound. It follows that as the 
palinstrophy grows to and reaches a maximum 
($d\langle|\nabla\xi|^2\rangle/dt\ge0$), the following inequality 
necessarily holds
\begin{eqnarray}
\label{nogrowth}
\langle|\Delta\xi|^2\rangle^{1/2} &\le& 
\frac{\norm\xi_\infty\langle|\xi|^2\rangle^{1/2}}{\nu}.
\end{eqnarray} 

Since $\langle|\nabla\xi|^2\rangle$ can be bounded from above in terms of 
$\langle|\Delta\xi|^2\rangle$, ineq. (\ref{nogrowth}) can be used to derive 
an explicit upper bound for the palinstrophy. By H\"{o}lder inequality 
one has$^{18}$
\begin{eqnarray}
\label{Holder}
\langle|\Delta\xi|^2\rangle &\ge& 
\frac{\langle|\nabla\xi|^2\rangle^2}{\langle|\xi|^2\rangle},
\end{eqnarray}
where the inequality sign ``$\ge$'' can become ``$\gg$'' (see below).
Substituting (\ref{Holder}) into (\ref{nogrowth}) yields
\begin{eqnarray}
\label{nogrowth0}
\langle|\nabla\xi|^2\rangle &\le& 
\frac{\norm\xi_\infty\langle|\xi|^2\rangle}{\nu}.
\end{eqnarray} 
It follows that the enstrophy dissipation $\chi$ is bounded from above by
\begin{eqnarray}
\label{chibound}
\chi &\le& \norm\xi_\infty\langle|\xi|^2\rangle.
\end{eqnarray}
It is notable that the upper bound for $\chi$ in (\ref{chibound}) is 
expressible in terms of two decaying Casimirs, namely $\norm\xi_\infty$ 
and $\langle|\xi|^2\rangle$, so that it can be bounded from above in terms 
of initial data of the flows. More accurately, $\chi$ can be bounded from 
above in terms of the initial vorticity field only. In passing, it is worth 
mentioning that since $\norm\xi_\infty$ and $\langle|\xi|^2\rangle$ are 
intensive quantities, i.e. independent of the domain size, the constraint 
(\ref{chibound}) is size-independent. 

Let us denote by $k_d$ and $k_D$ the wavenumbers defined by 
$\langle|\nabla\xi|^2\rangle^{1/2}/\langle|\xi|^2\rangle^{1/2}$ and 
$\langle|\Delta\xi|^2\rangle^{1/2}/\langle|\nabla\xi|^2\rangle^{1/2}$, 
respectively. For ``regular'' spectra, $k_d$ ($k_D$) specifies where, 
in wavenumber space, $\langle|\nabla\xi|^2\rangle$ 
($\langle|\Delta\xi|^2\rangle$) is mainly distributed. In other words, 
$k_d$ ($k_D$) specifies where, in wavenumber space, the enstrophy 
(palinstrophy) dissipation mainly occurs. By ``regular'' it is meant 
that the enstrophy is not highly concentrated in any particular regions 
of wavenumber space that would result in severe steps in the enstrophy 
spectrum. By (\ref{Holder}), these dissipation wavenumbers satisfy 
$k_d\le k_D$, and the inequality sign ``$\le$'' can become ``$\ll$''. 
This can be realized if the palinstrophy spectrum around $k_d$ does not 
fall off so steeply. More quantitatively, by the definition of $k_d$, one 
can expect the palinstrophy spectrum around $k_d$ to be shallower than 
$k^{-1}$. (Because, otherwise, most of the contribution to 
$\langle|\nabla\xi|^2\rangle$ would come from $k<k_d$, making the ratio 
$\langle|\nabla\xi|^2\rangle^{1/2}/\langle|\xi|^2\rangle^{1/2}$ 
significantly lower than $k_d$, a contradiction to the very definition 
of $k_d$.) This means that the spectrum of $\langle|\Delta\xi|^2\rangle$ 
around $k_d$ is shallower than $k^{1}$. Hence, most of the contribution to 
$\langle|\Delta\xi|^2\rangle$ can come from $k\gg k_d$ if the palinstrophy 
spectrum beyond $k_d$ becomes steeper than $k^{-1}$ and falls off to 
$k^{-3}$ gradually. This allows for the possibility $k_D\gg k_d$ to be 
realized. In any case, $k_d$ should be well beyond the end of the enstrophy 
range, i.e. $\int_{k>k_d}Z(k)\,dk/\int_{k<k_d}Z(k)\,dk\approx0$, and 
$k_D$ should be well beyond the palinstrophy range, i.e. 
$\int_{k>k_D}P(k)\,dk/\int_{k<k_D}P(k)\,dk\approx0$, 
where $Z(k)$ and $P(k)$ are the enstrophy and palinstrophy spectra, 
respectively. In other words, the enstrophy spectrum around $k_d$ should 
be steeper than $k^{-1}$, and the palinstrophy (enstrophy) spectrum around 
$k_D$ should be steeper than $k^{-1}$ ($k^{-3}$).

Our primary concern is an estimate of $k_d$ when $\langle|\nabla\xi|^2\rangle$ 
achieves a maximum. It is likely that $k_d$ achieves a global maximum 
then. Equation (\ref{Pevolution}) and the subsequent equations 
(\ref{nogrowth}) and (\ref{nogrowth0}) imply
\begin{eqnarray}
\label{kdkD}
k_dk_D &\le& \frac{\norm\xi_\infty}{\nu}
\end{eqnarray}
and
\begin{eqnarray}
\label{kd}
k_d &\le& \left(\frac{\norm\xi_\infty}{\nu}\right)^{1/2} 
\le \left(\frac{\norm{\xi(0)}_\infty}{\nu}\right)^{1/2}.
\end{eqnarray}

It is interesting to compare the present result with the classical 
dissipation wavenumber $k_\nu$. To this end, let us express 
$\chi=\nu\langle|\nabla\xi|^2\rangle$ in the form 
$\chi=\nu k_d^2\langle|\xi|^2\rangle$, so that (\ref{knu}) can be 
rewritten as
\begin{eqnarray}
\label{knu1}
k_\nu&=&\frac{\chi^{1/6}}{\nu^{1/2}}=
\left(\frac{k_d\langle|\xi|^2\rangle^{1/2}}{\nu}\right)^{1/3}.
\end{eqnarray}
It follows that
\begin{eqnarray}
\label{comparision}
\frac{k_\nu^3}{k_d} &=& \frac{\langle|\xi|^2\rangle^{1/2}}{\nu}.
\end{eqnarray}
It may be assumed that $\norm\xi_\infty$ and $\langle|\xi|^2\rangle^{1/2}$ 
are of the same order of magnitude. In fact, one can even have the exact 
equality $\norm\xi_\infty=\langle|\xi|^2\rangle^{1/2}$ for some simple cases. 
For example, for a vorticity field of a single Fourier mode, this equality 
trivially holds. In any case, both $\norm\xi_\infty$ and 
$\langle|\xi|^2\rangle^{1/2}$ decay under the action of viscosity, 
so that if they are comparable initially, then one can expect them to be 
comparable subsequently. Hence one can deduce from (\ref{kdkD}) and 
(\ref{comparision}) that
\begin{eqnarray}
\label{comparision1}
k_d^2k_D \approx k_\nu^3.
\end{eqnarray}
It follows that $k_d\le k_\nu$, and $k_d$ can be significantly lower 
than $k_\nu$ if $k_d\ll k_D$.

As pointed out by a referee, the result (\ref{kd}) can be derived from 
more physical considerations, based on the relation
\begin{eqnarray}
\label{referee}
\partial_t\langle\xi(1)\xi(2)\rangle +
\partial_r\langle\delta v \xi(1)\xi(2)\rangle &\approx&
\nu\partial_r^2\langle\xi(1)\xi(2)\rangle,
\end{eqnarray}
where 1 and 2 stand for $\bm x_1$ and $\bm x_2$, respectively, 
$r=|\bm x_2-\bm x_1|$, and $\delta v$ denotes the fluid speed increment 
across $r$. In the limit $r\rightarrow k_d^{-1}$, if one assumes 
$\xi\approx\delta v/r$ then one can recover (\ref{kd}) by 
balancing the nonlinear and viscous terms in (\ref{referee}). 

In conclusion, this letter has derived a rigorous constraint on the 
palinstrophy production rate by nonlinear transfer, from which upper 
bounds for the enstrophy dissipation $\chi$ and for the enstrophy 
dissipation wavenumber $k_d$ have been deduced. These bounds are expressible 
in terms of the vorticity supremum  $\norm\xi_\infty$, the mean-square 
vorticity density $\langle|\xi|^2\rangle$, and the 
viscosity $\nu$. These quantities are ``intensive'' quantities, i.e. 
independent of the domain size, making the derived upper bounds 
size-independent. Moreover, these bounds are completely determined by 
the initial vorticity field and $\nu$ since both $\norm\xi_\infty$ and 
$\langle|\xi|^2\rangle$ decay under the action of viscosity. The upper 
bound $k_d\le\norm{\xi(t)}_\infty^{1/2}/\nu^{1/2}\le\norm{\xi(0)}_\infty
^{1/2}/\nu^{1/2}$ is consistent with the classical dissipation wavenumber 
$k_\nu=\chi^{1/6}/\nu^{1/2}$, in the 
sense that they both have the same functional dependence on $\nu$. A 
novel result of this study is that both $\chi$ and $k_d$ are bounded in 
terms of the initial vorticity field and the viscosity whereas $\chi$ 
and $k_\nu$ are essentially undetermined by the classical theory.  

It is a pleasure to acknowledge stimulating discussions with Robert Kerr,
Sergey Nazarenko, and James Robinson during the course of this work. Djoko 
Wirosoetisno carefully read the manuscript and offered helpful comments.

~

\noindent $^1$G. K. Batchelor, ``Computation of the energy spectrum in 
homogeneous two-dimensional turbulence,'' Phys. Fluids {\bf 12}, 233 (1969).
 
\noindent $^2$G. Boffetta, A. Celani, and M. Vergassola, ``Inverse energy 
cascade in two-dimensional turbulence: Deviations from Gaussian behaviour,'' 
Phys. Rev. E {\bf 61}, R29 (2000).

\noindent $^3$V. Borue, ``Inverse energy cascade in stationary two-dimensional 
homogeneous turbulence,'' Phys. Rev. Lett. {\bf 72}, 1475 (1994); 
``Spectral exponents of enstrophy cascade in stationary two-dimensional 
homogeneous turbulence,'' Phys. Rev. Lett. {\bf 71}, 3967 (1993).

\noindent $^4$S. Chen {\it et al.}, ``Physical machanism of the 
two-dimensional enstrophy cascade,'' Phys. Rev. Lett. {\bf 91}, 214501 (2003).

\noindent $^5$P. Constantin, C. Foias, and O. P. Manley, ``Effects of the 
forcing function spectrum on the energy spectrum in 2-D turbulence,''
Phys. Fluids {\bf 6}, 427 (1994).

\noindent $^6$B. Dubrulle and S. Nazarenko, ``Interaction of turbulence
and large-scale vortices in incompressible 2D fluids,'' Physica D {\bf 110},
123 (1997).

\noindent $^7$G. Eyink, ``Exact results on stationary turbulence in 2d: 
consequences of vorticity conservation,'' Physica D {\bf 91}, 97 (1996).

\noindent $^8$R. Fj\o{}rtoft, ``On the changes in the spectral distribution 
of kinetic energy for twodimensional, nondivergent flow,'' Tellus {\bf 5}, 
225 (1953).

\noindent $^9$U. Frisch and P. L. Sulem, ``Numerical simulation of the inverse
cascade in two-dimensional turbulence,'' Phys. Fluids {\bf 27}, 1921 (1984).

\noindent $^{10}$R. H. Kraichnan, ``Inertial ranges in two-dimensional 
turbulence,'' Phys. Fluids {\bf 10}, 1417 (1967); ``Inertial-range transfer 
in two- and three-dimensional turbulence,'' J. Fluid Mech. {\bf 47}, 
525 (1971).

\noindent $^{11}$S. B. Kuksin, ``The Eulerian limit for 2D statistical 
hydrodynamics,'' J. Stat. Phys. {\bf 115}, 469 (2004).

\noindent $^{12}$C. E. Leith, ``Diffusion approximation for two-dimensional 
turbulence,'' Phys. Fluids {\bf 11}, 671 (1968).

\noindent $^{13}$E. Lindborg and K. Alvelius, ``The kinetic energy spectrum 
of the two-dimensional enstrophy turbulence cascade,'' Phys. Fluids {\bf 12}, 
945 (2000).

\noindent $^{14}$P. E. Merilees and H. Warn, ``On energy and enstrophy 
exchanges in two-dimensional non-divergent flow,'' J. Fluid Mech. {\bf 69}, 
625 (1975).

\noindent $^{15}$J. Paret, M.-C. Jullien, and P. Tabeling, ``Vorticity 
Statistics in the two-dimensional enstrophy cascade,'' Phys. Rev. Lett. 
{\bf 83}, 3418 (1999).

\noindent $^{16}$L. M. Smith, V. Yakhot, ``Bose condensation and small-scale 
structure generation in a random force driven 2D turbulence,'' Phys. Rev. 
Lett. {\bf 71}, 352 (1993); ``Finite-size effects in forced, two-dimensional 
turbulence,'' J. Fluid Mech. {\bf 271}, 115 (1994).

\noindent $^{17}$C. V. Tran and J. C. Bowman, ``Robustness of the inverse 
cascade in two-dimensional turbulence,'' Phys. Rev. E {\bf 69}, 036303 
(2004); ``On the dual cascade in two-dimensional turbulence,'' Physica D 
{\bf 176}, 242 (2003).

\noindent $^{18}$C. V. Tran, ``Nonlinear transfer and spectral distribution 
of energy in $\alpha$ turbulence,'' Physica D {\bf 191}, 137 (2004).

\noindent $^{19}$C. V. Tran and T. G. Shepherd, ``Constraints on the spectral 
distribution of energy and enstrophy dissipation in forced two-dimensional 
turbulence,'' Physica D {\bf 165}, 201 (2002).

\noindent $^{20}$D. D. Holm {\it et al.}, ``Nonlinear stability of fluid
and plasma equilibria,'' Phys. Rep. {\bf 123}, 1 (1985).

\noindent $^{21}$Y. C. Li, ``On 2D Euler equations. I. On the 
energy-Casimir stabilities and the spectra for linearized 2D Euler equations,''
J. Math. Phys. {\bf 41}, 728 (2000).

\noindent $^{22}$D. Wirosoetisno and T. G. Shepherd, ``Nonlinear stability
of Euler flows in two-dimensional periodic domains,'' Geophys. Astrophys. 
Fluid Dynamics {\bf 90}, 229 (1999).

\end{document}